\documentclass[graybox, nosecnum]{svmult}

\usepackage{mathptmx}       
\usepackage{helvet}         
\usepackage{courier}        
\usepackage{type1cm}        
\usepackage{makeidx}         
\usepackage{graphicx}        
\usepackage{multicol}        
\usepackage[bottom]{footmisc}
\usepackage{hyperref}        
\usepackage{soul}            
\hypersetup{colorlinks=true,urlcolor=blue}
\usepackage{amsmath}
\usepackage{amssymb}

\usepackage[square,numbers]{natbib}

\emergencystretch=\maxdimen
\makeindex             

\begin{document}
\title*{Detection Landscape in the Deci-Hertz Gravitational-Wave Spectrum}
\author{Kiwamu Izumi \thanks{corresponding author}, Karan Jani}

\institute{Kiwamu Izumi \at Institute of Space and Astronautical Science, Japan Aerospace Exploration Agency, 3-1-1 Yoshinodai, Chuo, Sagamihara, Kanagawa 252-5210 Japan, \email{kiwamu@astro.isas.jaxa.jp}, \\ \\
Karan Jani \at  Department Physics and Astronomy, Vanderbilt University, 2301 Vanderbilt Place, Nashville, TN, 37235, USA , \email{karan.jani@vanderbilt.edu}
}
%
%
\maketitle
\abstract{Direct observations of gravitational waves at frequencies around deci-Hertz will play a crucial role in fully exploiting the potential of multi-messenger astronomy. In this chapter, we discuss the detection landscape for the next several decades of the deci-Hertz gravitational-wave spectrum. We provide an overview of the experimental frontiers being considered to probe this challenging regime and the astrophysics and fundamental goals accessible towards them. This includes interferometric observatories in space with heliocentric and geocentric satellites, cubesats, lunar-based experiments and atom intereferometry. A major focus of this chapter is towards the technology behind DECi-hertz Interferometer Gravitational wave Observatory (DECIGO) and its scientific pathfinder mission concept B-DECIGO.  }
\section{Keywords} 
Instruments, laser interferometry, space mission, lunar science, cubesats, atomic clock, gravitational-wave astronomy.

\section{1. Introduction}
The current generation of ground-based gravitational-wave detectors such as Laser Interferometer Gravitational-Wave Observatory (LIGO) \cite{TheLIGOScientific:2014jea}, Virgo \cite{TheVirgo:2014hva} and KAGRA \cite{Akutsu:2018axf} are sensitive at frequencies above 10~Hz. These L-shaped detectors with a few km arm-length are suited to probe gravitational waves from the coalescing binaries of neutron stars and stellar black holes $({\lesssim}10^2~M_\odot)$ \cite{Aasi:2013wya}. The next-generation of ground-based gravitational-wave detectors like Einstein Telescope \cite{Punturo:2010zz} and Cosmic Explorer \cite{Reitze:2019iox} are expected to be sensitive as low as 5~Hz. The arm-length of these detectors will be tens of km and include several technological upgrades from the current generation. With these next-generation detectors, the sensitive volume for stellar compact binaries will reach unprecedented cosmological scale, and open the prospects for direct detections of lower-range to mid-range intermediate-mass black hole $({\lesssim}10^3~M_\odot)$ \cite{Jani:2019ffg}. The upcoming space-based gravitational-wave detector Laser Interferometer Space Antenna (LISA) \cite{Audley:2017drz} will have a few million km long arm length, thus measuring low-frequency gravitational waves near the milli-Hertz range. This debut space-based gravitational-wave detector will open a new window into the astrophysics of super-massive black hole binaries of ${\sim}10^6-10^7~M_\odot$. Furthermore, the ongoing network of pulsar timing arrays \cite{IPTA):2013lea} are probing the nano-Hertz regime. This can potentially measure gravitational waves from super-massive black hole binaries as heavy as ${\sim}10^9~M_\odot$ \cite{Mingarelli:2019pyd}. 

A particularly challenging regime to measure gravitational waves for all the detectors mentioned above is near the deci-Hertz frequencies $({\sim}0.1~\mathrm{Hz})$. For a terrestrial laser interferometric setup, it is difficult to go below 1~Hz barrier due to earth's seismic noise. On the other hand, space missions like LISA cannot go higher than 0.1~Hz due to laser shot noise. The universe is filled with rich astrophysical sources in and around deci-Hertz $(10^{-2}~\mathrm{Hz}\sim1~\mathrm{Hz})$ (see \cite{Sedda:2019uro, Mandel:2017pzd} for a recent review). The chief among them are the elusive intermediate-mass black holes (IMBHs: $10^{2}-10^5~M_\odot$), which are crucial clues in understanding the first population of stars in the universe, as well as the seed formation of the super-massive black holes found at the centers of galaxies (see \cite{Greene_IMBH, Bellovary:2019nib} for a recent review). There are a few promising candidates for IMBHs across its broad mass-range from electromagnetic observations, though the first confirmed evidence is attributed to the remnant of binary black hole merger GW190521, observed by the LIGO and Virgo detectors \cite{Abbott_2020_190521}. 


\begin{figure}[t]
\centering
	\includegraphics[scale=0.2, trim = {100 0 0 0}]{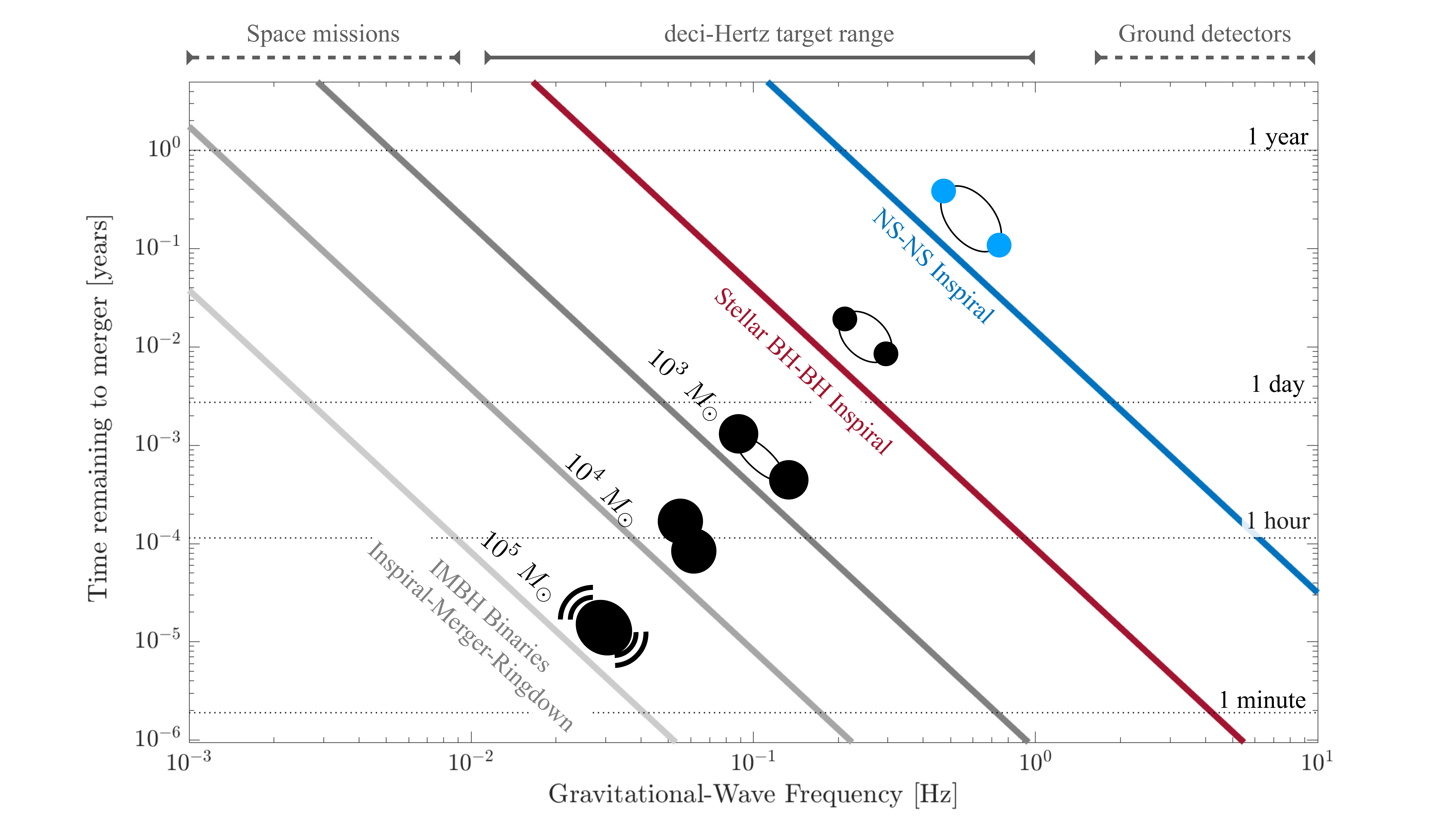}
	\caption{Gravitational-wave frequencies corresponding to the evolution of some of the prominent astrophysical binaries in the deci-Hertz spectrum. The inspiral of binary neutron star (blue line) and stellar binary black holes (maroon line) will be observed in the deci-Hertz for years. For intermediate-mass black holes, IMBHs, (gray lines), depending on their mass-ranges, the deci-Hertz access would measure their inspiral, merger and ringdown. }
	\label{fig:merge}
\end{figure}

For a coalescing binary, the instantaneous gravitational-wave frequency $f_\mathrm{GW}$ at a time $\tau$ before the merger scales inversely with the chirp mass $\mathcal{M}$ \cite{Peters:1964zz}. In Fig. \ref{fig:merge}, we show the evolution of various astrophysical binaries around the deci-Hz range. Binaries with IMBHs would generally undergo all the three phases of the coalescence - inspiral, merger and ringdown - within in the deci-Hertz band. A binary with an upper-range $({\sim}10^5~M_\odot)$ and mid-range $({\sim}10^4~M_\odot)$ IMBHs would undergo inspiral for a few minutes to hours in the deci-Hertz band, followed by merger and ringdown. Binaries with lower-range $({\sim}10^3~M_\odot)$ IMBHs would inspiral in deci-Hertz spectrum for a few days, though their merger and ringdown may be better accessible with next-generation ground-based detectors. The expected binary population of IMBHs remain fairly unknown and the constraints on their merger rates differ dramatically from equal-mass systems to intermediate-mass ratios $(m_1/m_2 \sim 100)$. The deci-Hertz spectrum opens a rare observational window into testing a variety of formation scenarios of IMBHs to cosmological distances. 

Within the frequency band of the current generation detectors such as LIGO, Virgo and KAGRA $(f_\mathrm{GW} = [10,~1000]~\mathrm{Hz})$, we can only measure about a second of evolution of binaries of stellar black holes $(\mathcal{M}\sim 10~M_\odot)$. In the deci-Hertz band, these binary black holes stay for months (maroon line). The binary neutron star sources would evolve within deci-Hertz band up to a few years (blue line). This extended observation is important to measure the orbital eccentricity and spin orientation of the binaries. Both of these measurements are key to distinguish among binary formation channels \cite{Stephan:2016kwj, Rodriguez:2016vmx}. A deci-Hertz range detector also acts as an early-alert system. All the binaries that will be seen by next-generation ground-based detectors such as Einstein Telescope and Cosmic Explorer, would be seen at the lower frequency many months ahead by a deci-Hertz detector. This opens an interesting possibility of multiband gravitational-wave astronomy, wherein the same astrophysical source is studied in different frequency bands \cite{Jani:2019ffg, Sesana:2016ljz, Cutler:2019krq, Isoyama:2018rjb}. An equivalent analogy of such in astronomy is that of multi-wavelength electromagnetic observations across radio, X-ray and other bands. The precise knowledge of sky-localization and distance from deci-Hertz observation will permit high-latency electromagnetic followups and also stronger constraints on the Hubble constant with ``dark sirens'' \cite{GLOC}.

The deci-Hertz spectrum allows a unique peak into the progenitors of Type Ia supernovae (such as double white dwarfs \cite{Toonen:2012jj}), whose observations will help for better calibration of standard candles \cite{GLOC, Lunar_Harms}. The low-frequency gravitational waves by the asymmetric ejection of neutrino during core-collapse supernovae will also be in the deci-Hertz regime \cite{Vartanyan:2020nmt}. In context of fundamental physics, the deci-Hertz sources open a wide range of tests for general relativity, including any presence of scalar fields and dipole radiation. The multiband observations between deci-Hertz with LIGO-like ground detectors and LISA-like space missions will allow an independent check on the pre- and post-merger properties of black holes, thus aiding the tests of the no-hair theorem. For probing Beyond Standard Model physics, deci-Hertz will particularly be effective for probing bosonic fields around intermediate-mass black holes merger, sub-solar dark matter candidates and stochastic gravitational-wave background from electroweak symmetry breaking (see \cite{Sedda:2019uro} for a recent review). 

In this chapter, we will discuss the various experimental frontiers that are being considered to access this challenging yet rich deci-Hertz gravitational-wave spectrum. We will briefly discuss the astrophysics and fundamental physics goals that can be advanced with each of these frontiers, and comment on their feasibility by the next decade. A major part of this chapter focuses on the Japanese space-mission, DECIGO \cite{Kawamura:2011zz}, and its planned technological development.   


\section{Experimental Frontiers}
\label{Sec:frontiers}

\begin{figure}[t]
\centering
	\includegraphics[scale=0.18, trim = {60 1200 40 0}]{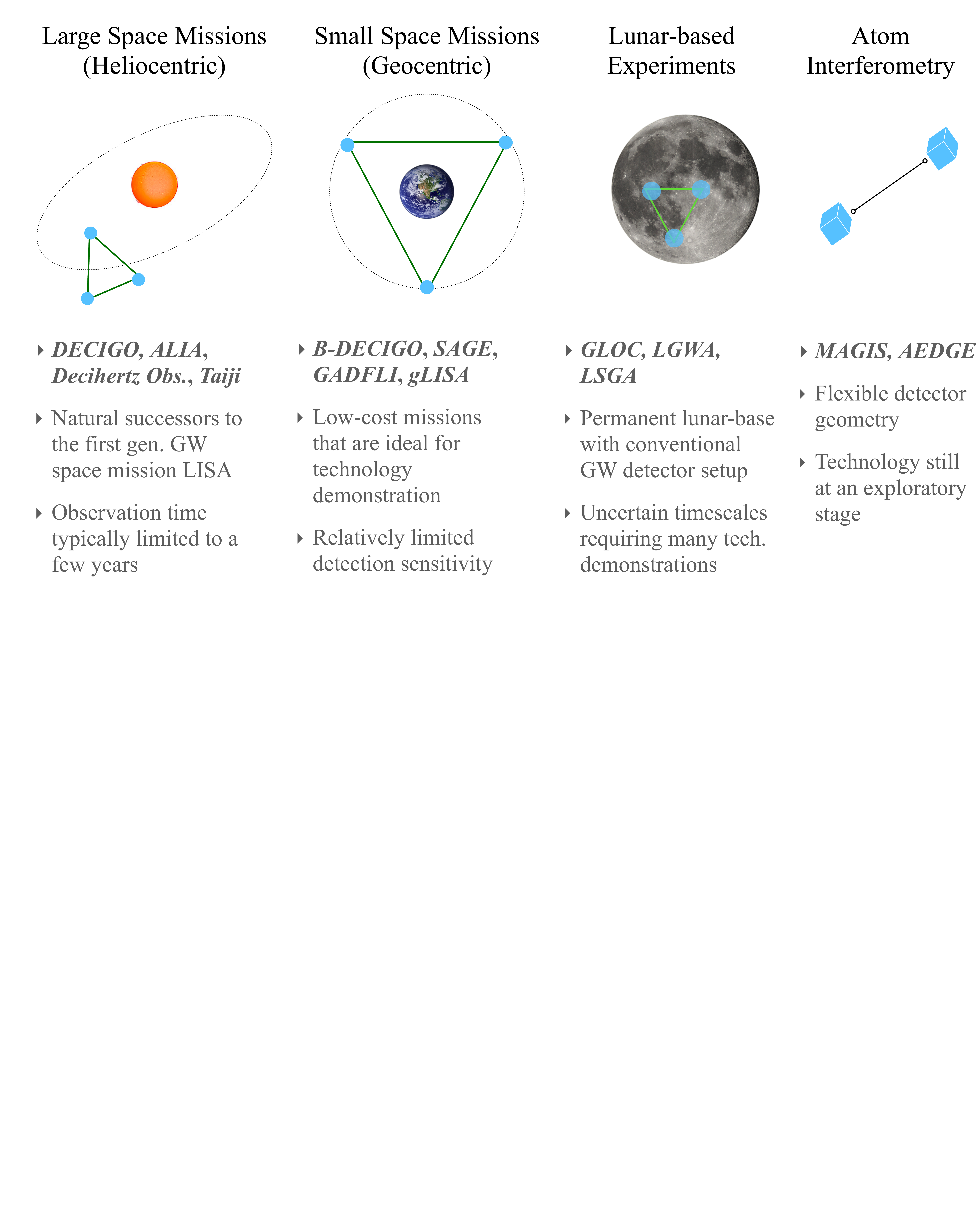}
	\caption{The four broad categories of experimental frontiers being considered for measuring gravitational-wave frequencies in the deci-Hertz regime. For each category, we list the prominent detector designs and their salient features. }
	\label{fig:chart}
\end{figure}

For an ideal deci-Hertz gravitational-wave detector, we expect its peak sensitivity around ${\sim}0.1$~Hz to be somewhere between that of LISA near the milli-Hz range (characteristic strain, $h=\delta{L}/L\sim10^{-21}$) or that of LIGO around 100~Hz ($h\sim10^{-22}$). While that certainly is not a strict criteria, it sets a target sensitivity for detecting expected gravitational-wave sources. Fig. \ref{fig:chart} highlights the four broad categories of experimental frontiers that are being considered to achieve the desired sensitivity in and around the deci-Hertz regime. This includes a space-based mission with (i) heliocentric or (ii) geocentric orbits, (iii) an experiment on the lunar surface, and (iv) atomic interferometry. Under each category, we list some of the prominent detector concepts suggested in the literature, along with their notable advantages and challenges. 

\begin{figure}[t]
\centering
	\includegraphics[scale=0.24, trim = {250 0 0 50}]{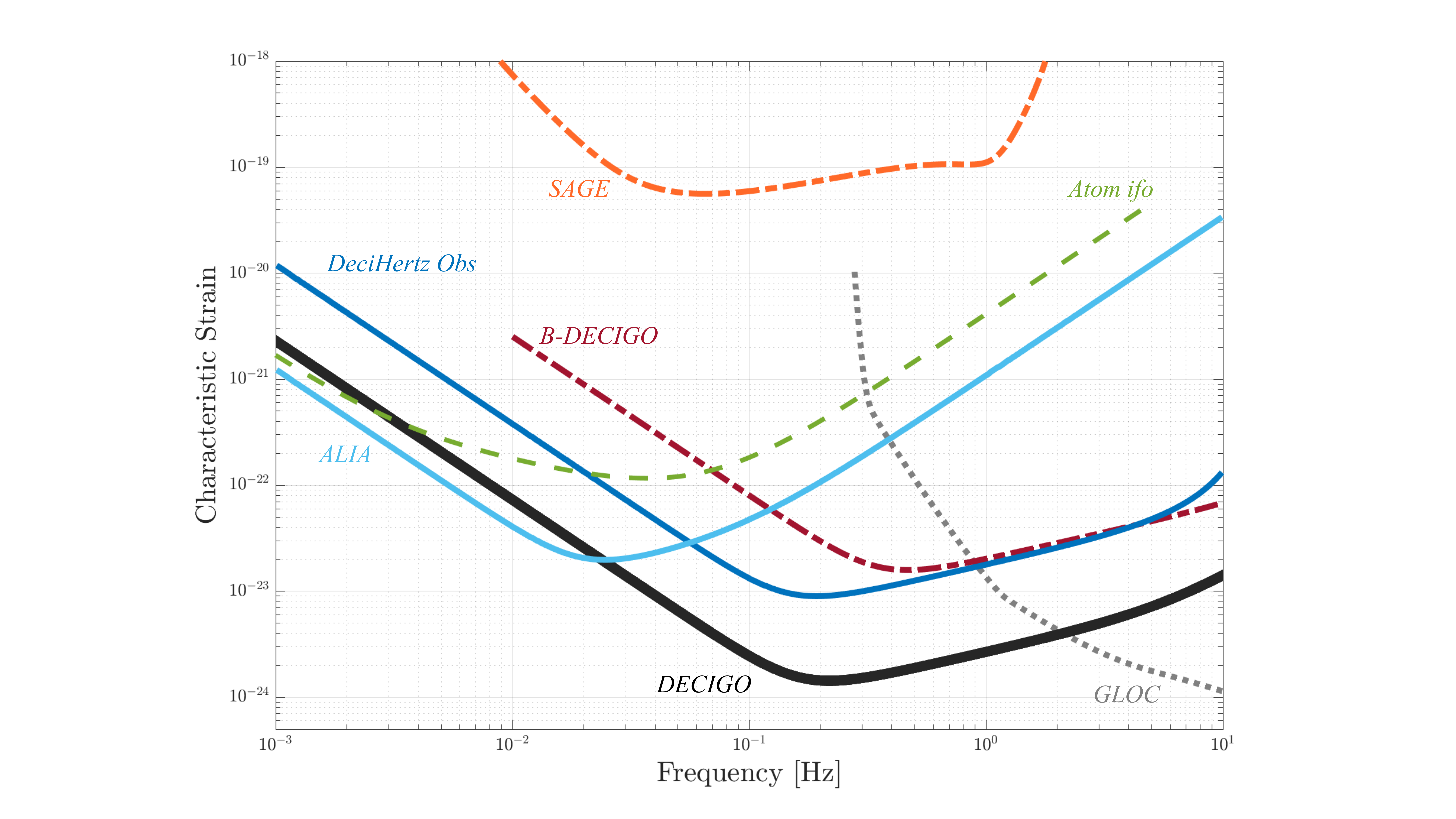}
	\caption{Sensitivity of the proposed gravitational-wave experiments that can measure frequencies near the deci-Hz spectrum (0.1 Hz). The thick black line refers to the most sensitive heliocentric space-mission in this region, DECIGO. The maroon dotted-dashed line refers to its pathfinder space-mission in geocentric orbit, B-DECIGO. Other heliocentric space missions shown are ALIA (sky blue curve) and DeciHertz Observatory (blue curve). The lunar-based detector, GLOC, is shown with dotted grey lines. More unconventional approaches such as  cubesat mission SAGE (orange dotted-dashed curve) and a detector based on atomic interferometer (dashed green lines) are also highlighted.        }
	\label{fig:sense}
\end{figure}

\begin{figure}[t]
\centering
	\includegraphics[scale=0.21, trim = {180 0 0 0}]{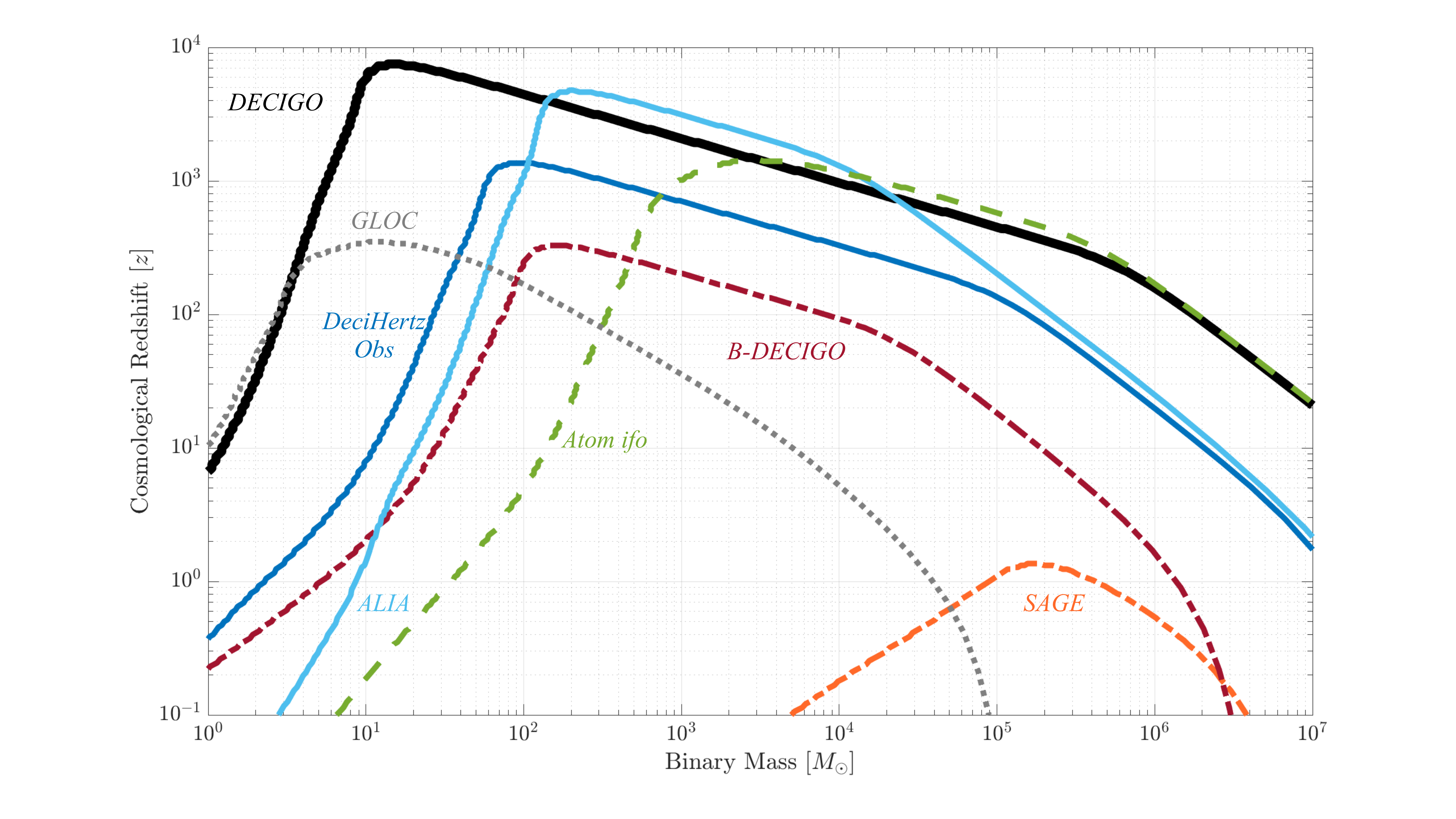}
	\caption{The Detection radius measured in the units of cosmological redshift $z$ for the proposed deci-Hertz gravitational-wave experiments. The colors refer to the detector sensitivities shown in Fig. \ref{fig:sense}. The y-axis is equivalent to the horizon distance for an optimally placed, equal-mass, non-spinning binary. We compute the horizon for SNR=8. In the case of space-based detectors (all detectors except GLOC), we put a detector lifetime of 5 years. The x-axis refers to the total mass of the binary as measured in its source frame, i.e. with the redshift correction $1/(1+z)$. We compute redshift assuming the Planck 2018 cosmological parameters. }
	\label{fig:vol}
\end{figure}

We plot the sensitivity curves for some of the proposed deci-Hertz detectors in Fig. \ref{fig:sense}. The detector sensitivity is measured as the dimensionless characteristic strain $h = \delta{L}/L$. The line styles refers to the four categories of experimental frontiers: heliocentric (straight lines), geocentric (dotted-dashed lines), lunar-based (dotted line) and atom interferometry (dashed line). In Fig. \ref{fig:vol}, we show the horizon distance, i.e. the maximum detection radius, of each of these detectors towards coalescing binaries. Notice that other than the cubesat mission SAGE, all the proposed deci-Hertz detector can reach up to redshift $z\sim100$ for a certain mass-range of binaries.

A majority of the proposed efforts are geared towards a large-scale heliocentric space mission similar to that of LISA. With three satellites, each separated by 2.5 million kms arm-length forming the interferometer, the LISA mission can, in-principle, access gravitational waves with frequencies as low as ${\sim}0.1$~mHz and as high as ${\sim}0.1$~Hz. At this higher frequency end, LISA's sensitivity is expected to be at most $h\sim10^{-20}$. A post-LISA mission with a similar 3-satellite configuration but an overall higher sensitivity could achieve the target for deci-Hertz observations. Few of such design ideas include the Advanced Millihertz Gravitational-wave Observatory ({\it AMIGO})~\cite{Baibhav:2019rsa}, which was proposed to ESA's Voyage 2050 call, and the Chinese-based mission \textit{Taiji}~\cite{Hu:2017mde}.

To achieve the most of the deci-Hertz science, a detector should have high-sensitivity across a broad frequency coverage of $0.01-1$ Hz. For space-missions with LISA-like concepts, it will require shorter arm-length and technologies to reduce the acceleration noise. Two of such prominent design ideas are the Advanced Laser Interferometer Antenna ({\it ALIA})~\cite{Bender:2013nsa,Baker:2019pnp} and DeciHertz Observatories ({\it DO})~\cite{Sedda:2019uro}, proposed to ESA's Voyage 2050 call as well. Their sensitivities and detection radii are shown in Fig. \ref{fig:sense} and Fig. \ref{fig:vol} (sky-blue and dark-blue curves), respectively. Both these missions could detect IMBH binaries across its three orders of mass-range $(10^2-10^5~M_\odot)$ to an unprecedented cosmological distance (redshift $z\sim100$). Furthermore, they would also detect stellar $({\sim}10~M_\odot)$ and supermassive $({\sim}10^6~M_\odot)$ binary black holes binaries to $z\lesssim 10$, thus complementing with the science goals from LISA and LIGO.


More recently, there has been modest proposals for relatively low-cast space-missions in geocentric orbits that can access deci-Hertz region, albeit not with the same sensitivity requirement. This includes a triangular interferometeric constellation with cubesats, called {SagnAc interferometer for Gravitational wavE} (\textit{SAGE})~\cite{Lacour:2018nws}. The sensitivity and detection radius for SAGE is shown in Fig. \ref{fig:sense} and Fig. \ref{fig:vol} (orange dashed-dotted curve). The detection sensitivity of SAGE allows a survey of upper-range IMBH binaries $({\sim}10^5~M_\odot)$ within redshift $z\lesssim1$. Other proposals of such geocentric mission in the literature includes geosynchronous Laser Interferometer Space Antenna (\textit{gLISA})~\cite{Tinto:2011nr} and the Geostationary Antenna for Disturbance-Free Laser Interferometry (\textit{GADFLI})~\cite{McWilliams:2011ra}.

With the advent of the return to Moon, it is now finally possible to think of a gravitational-wave detector on the surface of the Moon that would take advantage of the natural conditions. A few recent proposals that are being studied to access deci-Hertz regime from the Moon are the Gravitational-wave Lunar Observatory for Cosmology (\textit{GLOC})~\cite{GLOC}, Lunar Gravitational-Wave Antenna (\textit{LGWA})~\cite{Lunar_Harms} and Lunar Seismic Gravitational wave Antenna (\textit{LSGA})~\cite{LSGA}. We discuss these proposals in more details in the later sections. A promising detector technology is also emerging from atom-based interferometry, which could access gravitational-waves in 0.01-1 Hz of frequency range. Some of the prominent design proposals are Mid-band Atomic Gravitational Wave Interferometric Sensor (\textit{MAGIS})~\cite{Graham:2017pmn} and the {Atomic Experiment for Dark Matter and Gravity Exploration in Space} (\textit{AEDGE})~\cite{Bertoldi:2019tck}. The sensitivity and detection radius for a concept atom interferometer (adapted from \cite{Sedda:2019uro}, with parameters based on \cite{Kolkowitz:2016wyg}) is shown in Fig. \ref{fig:sense} and Fig. \ref{fig:vol} (green dashed-curve). For more details regarding the science goals and design proposals for atom-interferometer, see the chapter ``Quantum sensors with matter waves for GW observation'' in this edition by Bertoldi, Bouyer and Canuel.

The most ideal deci-Hertz detector, which has been extensively studied in the literature, is the proposed Japanese space-mission DECi-hertz Interferometer Gravitational wave Observatory ({\it DECIGO}). The sensitivity and detection radius of DECIGO is shown in Fig. \ref{fig:sense} and Fig. \ref{fig:vol} (black curve). The detector posses a rare advantage of accessing gravitational-waves to cosmological distances from stellar to IMBHs to supermassive black holes. The mission design and the associated technologies for DECIGO are discussed in the next section.



\section{DECIGO}

\begin{figure}[t]
\centering
	\includegraphics[width=0.6\columnwidth]{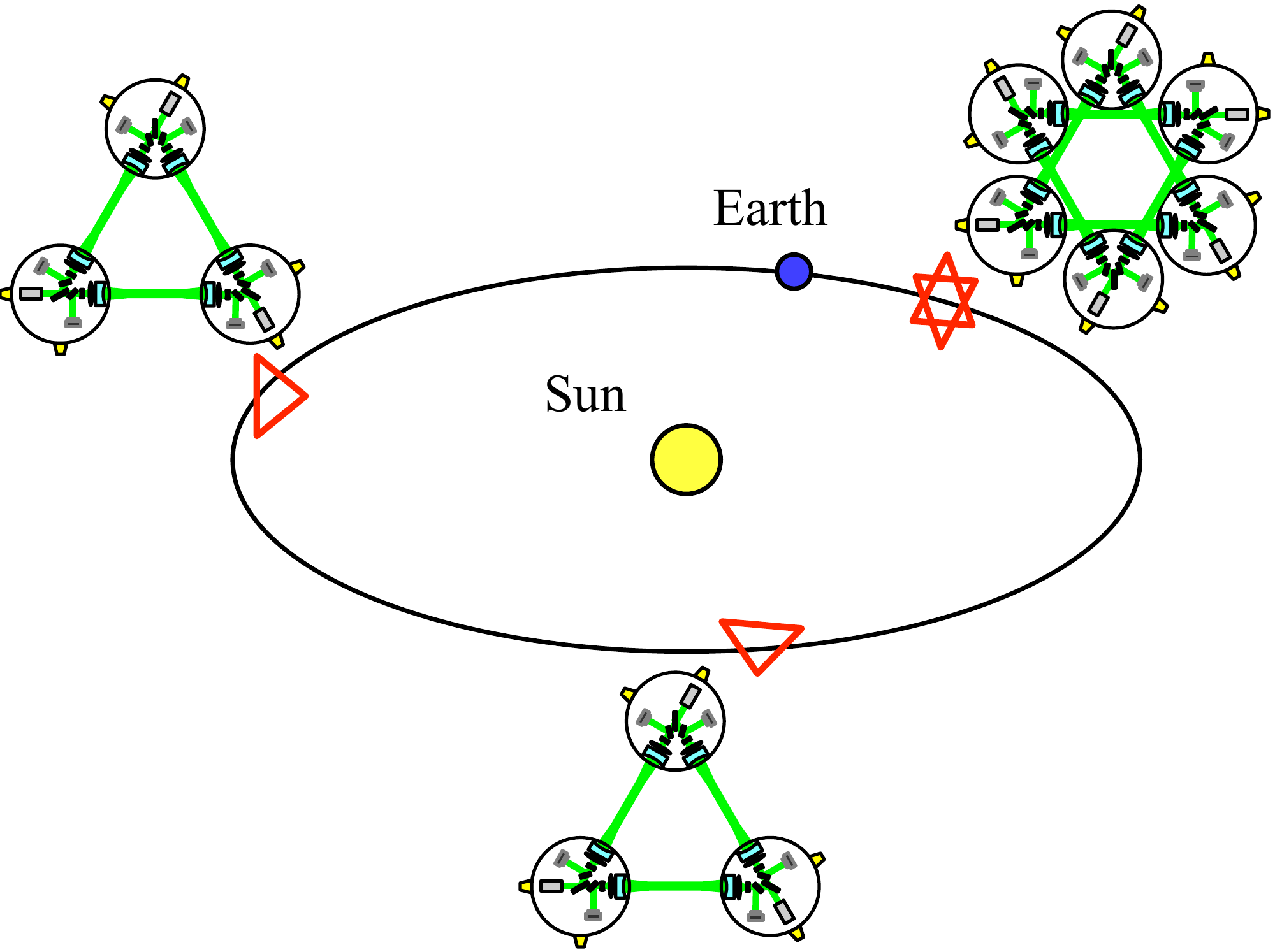}
	\caption{Orbit of DECIGO. Four clusters of observatories are distributed in the heliocentric orbit with two of them placed at the same position.}
	\label{fig:cluster}
\end{figure}

\begin{figure}[t]
\centering
\includegraphics[width=0.6\columnwidth]{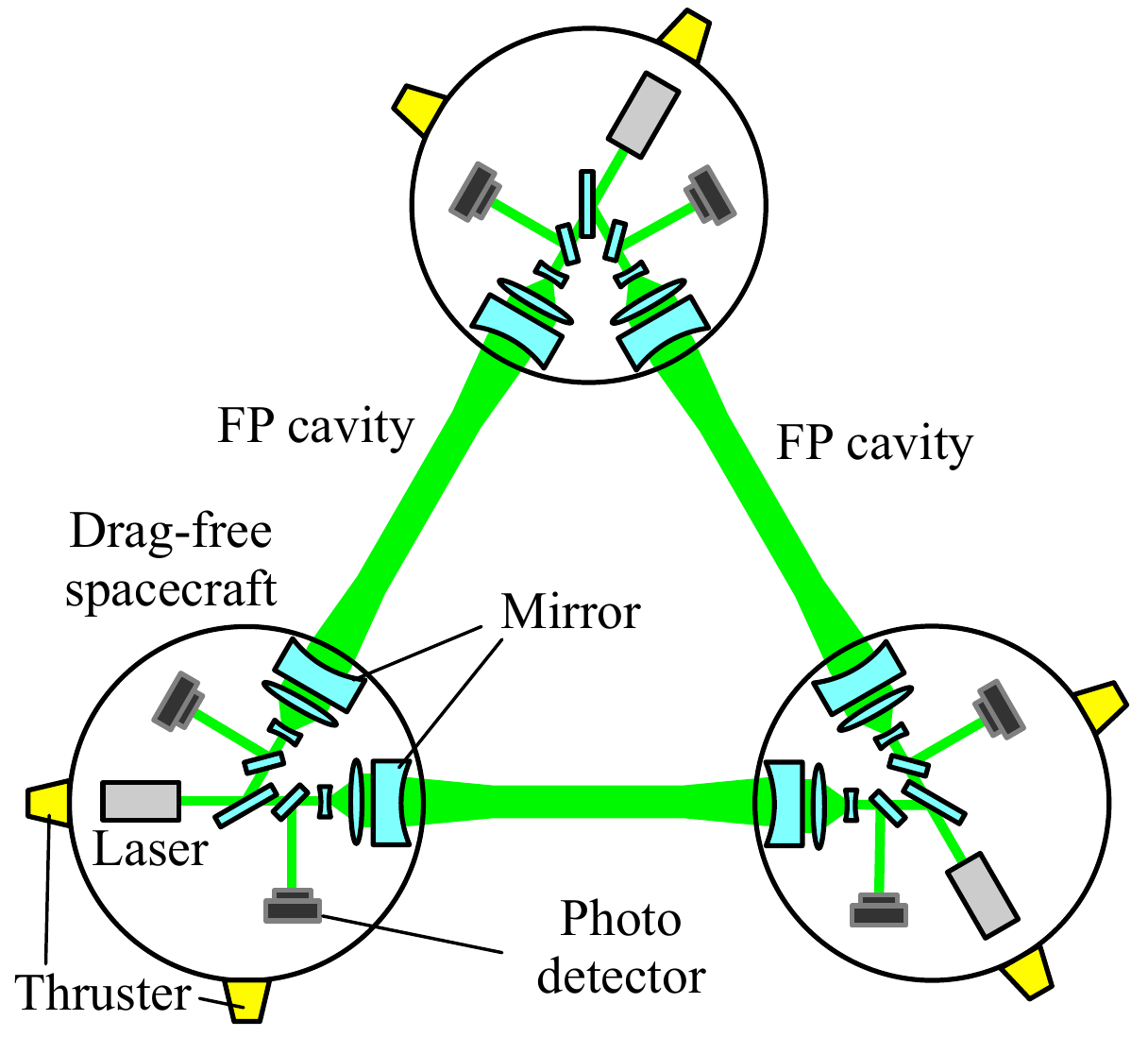}
\caption{Conceptual design of the interferometer for a single DECIGO cluster.}
\label{fig:design}
\end{figure}

DECi-hertz Interferometer Gravitational wave Observatory (DECIGO) is the future Japanese space gravitational wave antenna concept observing the frequency band from $\sim 10^{-1}$ to 10~Hz. DECIGO was originally proposed in 2001~\cite{Seto:2001qf} to fill the frequency gap between the mHz frequency band covered by LISA and the audio band by the terrestrial interferometers. The conceptual mission design calls for four identical clusters of observatories deployed in heliocentric orbits, as sketched in figure~\ref{fig:cluster}. Each of the clusters consists of three drag-free spacecraft, 1000~km apart from each other in almost equilateral triangular formation, forming six laser links~\cite{Kawamura_2006}. Each spacecraft houses two floating test masses serving as proof masses (see Figure~\ref{fig:design}). The target sensitivity of a single cluster is set to $4\times 10^{-24}\, \textrm{Hz}^{-1/2}$ at around 1~Hz~\cite{Kawamura:2020pcg}.

A unique design choice in DECIGO is the utilization of the Fabry-Perot optical cavities consisting of a pair of two test masses apart by 1000~km to improve the sensitivity without demanding a significant laser power. However, this means that the optical lengths of the Fabry-Perot cavities must be kept at resonances at a precision level of $10^{-9}$~m or less. Therefore, achieving a resonance of a Fabry-Perot cavity in orbit has been identified as one of the key challenges. The characteristic design parameters are summarized in table~\ref{tab:summary}.

\begin{table}[b]
\begin{center}
\begin{tabular}{l r r}
\hline \hline
 \textbf {Design Parameters }& \textbf{ DECIGO} & \textbf {B-DECIGO}\\ 
\hline
Laser wavelength & 515~nm & 515~nm\\
Laser power & 10~W & 1~W\\
Arm Length & 1000~km & 100~km\\
Finesse of the Fabry-Perot cavities& 10 & 100\\
Diameter of mirror (test mass) & 1~m & 30~cm\\
Mirror mass & 100~kg & 30~kg\\
Number of observatory clusters & 4 & 1\\
\hline
\end{tabular}
\caption{Design parameters for DECIGO and B-DECIGO~\cite{doi:10.1142/S0218271818450013}.\label{tab:summary}}
\end{center}
\end{table}

One of the main scientific objectives of DECIGO is to conduct searches for primordial gravitational wave backgrounds such as those amplified during the inflation period~\cite{Grishchuk_1993,PhysRevD.79.103501}. Besides, DECIGO will be capable of delivering new insight into astrophysics and fundamental physics through the observations of various compact star binary systems~\cite{Seto:2001qf,10.1143/PTP.123.1069} and other sources~\cite{PhysRevD.81.104043,PhysRevD.75.061302,Kakizaki:2015wua,PhysRevLett.124.041804}. In addition, heterogeneous observations of merger events with other gravitational wave detectors in different frequency bands should improve the performance of the astrophysical parameter estimation.

\subsection{B-DECIGO and technology developments}

In order to demonstrate the technologies relevant to DECIGO and simultaneously conduct astrophysically important observations in the deci-Hz band, the pathfinder mission concept called B-DECIGO was recently proposed~\cite{10.1093/ptep/ptw127}. The sensitivity and detection radius of B-DECIGO are shown in Fig. \ref{fig:sense} and Fig. \ref{fig:vol} (maroon dashed-dotted line). One of the major design drivers for B-DECIGO was the observatory performance being able to observe the stellar mass binary black holes up to a redshift of $z\sim 30$. The main scientific objective is to study the origin of the stellar mass black holes by mapping out their mass spectrum and event rate as a function of redshift. B-DECIGO is also capable of contributing to multi-messenger astronomy by issuing the forecasts for merger events with a sky localization error of $\sim 0.3 \,\textrm{deg}^2$ and a merger time accuracy of $\sim 1$~sec for a binary black hole with the masses similar to GW150914 at $z=0.1$, a day before the merger.

Based on the scientific motivations as well as the consideration for making the developments less challenging, B-DECIGO is designed to have a shorter arm length of 100~km and reduced mass of 30~kg for the test masses with only a single observatory cluster deployed. Geocentric orbits with a high altitude of 2000~km has been considered. The design parameters of B-DECIGO are also listed in table~\ref{tab:summary}. The target sensitivity is set to $2\times 10^{-23}\,\textrm{Hz}^{-1/2}$ at around 1~Hz. The launch date is currently foreseen to be some time in 2030's. 

The recent activities for the technology developments relevant to both DECIGO and B-DECIGO encompass a number of key functionalities and aspects. These include the laser source and its stabilization systems, the relative angular sensor for spacecraft~\cite{10.1117/12.2536073}, the interferometer control schemes~\cite{nagano2020demonstration, izumi2020backlinked}, the orbital design and further improvement in the sensitivity~\cite{YAMADA2020126626}.

\section{Lunar-Based Experiments}

The Moon offers a generous environment for gravitational-wave observations. There is no atmosphere, thus providing a high-quality vacuum just above the lunar surface. There are no ocean tides or significant ground motion to impact the differential displacement of the test masses. There are no large-scale human activities interfering with the detector. Therefore, in principle, one one can construct a large-scale interferometer like that of LIGO on lunar-soil. The conceptual idea has been long around \cite{Bender_Lunar}, though the halt of crewed missions to Moon since 70s dissipated any further interest from the community. It is only recently that initiatives such as NASA Artemis, Commercial Crew and ESA’s European Large Logistics Lander project that there is a strong possibility of returning to the Moon this decade and building a permanent base. One of the science priorities for NASA Artemis is in utilizing the unique environment of Moon to study the universe. 

Unlike space-based deci-Hertz detectors, a lunar-based detector can be in observational mode for decades. This is particularly important in following up rarer astrophysical processes in the deci-Hertz spectrum. Hardware failures or upgrading detector technologies is also possible on the Moon, thus providing a better value for long-term investment. Below we discuss two distinct experimental techniques for lunar-based gravitational-wave detection, which in particular opens the access to deci-Hertz gravitational-wave frequencies.  

\begin{figure}[t]
\centering
	\includegraphics[scale=0.27, trim = {0 0 0 0}]{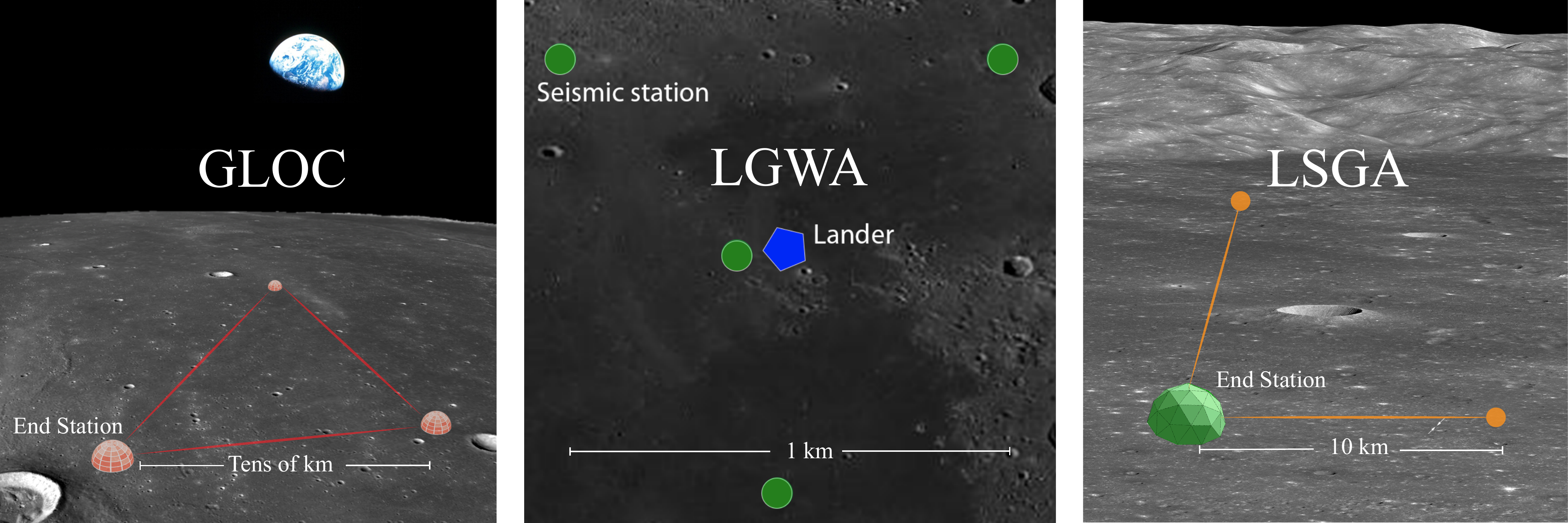}
	\caption{Conceptual designs of gravitational-wave detector on the lunar surface. Left: Gravitational-wave Lunar Observatory (GLOC) \cite{GLOC}. Right: Lunar Gravitational-Wave Antenna (LGWA) \cite{Lunar_Harms}. Figures have been reproduced by permission of the corresponding authors. }
	\label{fig:moon}
\end{figure}

The first technique is adapted for Gravitational-Wave Lunar Observatory (\textit{GLOC}), a proposed interferometer on the Moon \cite{GLOC}. The conceptual design for the GLOC is expected to be similar to that of the next-generation earth-based detectors Cosmic Explorer (CE) and Einstein Telescope (ET). The arm-length of the interferometer can be set to few tens of kms and the L-shaped LIGIO-like interferometer could be replaced by a triangular geometry such as that of ET (see Fig. \ref{fig:moon} - left panel). The end station could be covered in a temperature controlled dorm to host the required optics. Ideal sites for GLOC could be inside a crater or within one of lava caves. The tentative sensitivity curve and detection radius of GLOC is shown in Fig. \ref{fig:sense} and Fig. \ref{fig:vol} (gray dotted line). The lowest frequency we expect for GLOC is around ${\sim}0.3$ Hz. Below that frequency, the sensitivity of GLOC is limited by the seismic and thermal noises. However, compared to other deci-Hertz detectors, GLOC would remain sensitive at higher frequencies, thus enhancing the overall signal-to-noise ratio.

For stellar mass binaries, GLOC could detect them practically across the observable universe. The geocentric orbit of Moon helps tremendously in constraining the sky-localization to few arc-seconds in GLOC, thus opening new dark sirens for measuring Hubble constant. GLOC could measure sub-solar dark matter candidates upto redshift $z\sim10$. For IMBH binaries, GLOC could survey lower- to mid-range $({\sim}10^3-10^4~M_\odot)$ upto redshift $z\sim10$, while can find the upper-range $({\sim}10^5~M_\odot)$ within redshift $z\lesssim 1$. The rare access to ${\sim}1$~Hz in GLOC further opens avenues to test progenitor mechanisms to Type Ia, thus providing new calibrations for the standard candles.

A second technique has been shown in the Lunar Gravitational-Wave Antenna (\textit{LGWA}) \cite{Lunar_Harms} and the Lunar Seismic and Gravitational-Wave Antenna (\textit{LSGA}) \cite{LSGA}. An extension of the resonant-bar detector idea by Weber \cite{PhysRev.117.306}, in these concept designs the entire Moon acts as a large spherical detector receiving the incoming gravitational waves. The measurement of the signal is then essentially the vibrational eigenmodes. For LGWA, the proposal is to place an array of highend seismometers in km long array on the Moon (see Fig. \ref{fig:moon}-right panel). They will monitor the normal modes of the Moon in the frequency spectrum of 1~mHz — 1~Hz. The predicted sensitivity of LGWA near the deci-Hertz regime would be better than that of LISA, thus serving an important science driver. For LSGA, the proposal is to deploy 10 km optical cables in an L-shaped geometry. The laser light passing through the cables would permit the measurement of the displacement strain from incoming gravitational waves. Both these designs will be able to achieve the science goals of deci-Hertz gravitational-wave astronomy discussed earlier in the chapter.    

\section{Summary}

Continuing experimental effort have been seen over the world for designing gravitational-wave detectors sensitive at the frequencies around deci-Hertz. Observations in the deci-Hertz band will provide us with unique opportunities to enhance the ability of the detector network and to study the dark and relativistic aspects of our Universe. Among the large-scale space-missions, DECIGO, its scientific pathfinder mission concept B-DECIGO, ALIA and DO are expected to open the new gravitational window in the deci-Hertz band up to cosmological distances. For modest geocentric proposals, SAGE will be able to survey the deci-Hertz spectrum in the local universe. The lunar-based experiments GLOC, LGWA and LSGA offers a new avenue to measure frequencies near deci-Hertz, and substantially advance the scope of gravitational-wave astronomy.  

\section{Cross-Reference}
Bertoldi A., Bouyer P. and Canuel, B., ``Quantum sensors with matter waves for GW observation''

\bibliographystyle{iopart-num}
\providecommand{\newblock}{}

\end{document}